\documentclass[a4paper,12pt]{article}
\usepackage{amsmath, epsfig,graphicx}

\title{Hedging LIBOR Derivatives in a Field Theory Model of Interest Rates}
\author{Belal E. Baaquie, Cui Liang\\
Department of Physics, National University of Singapore\\
Kent Ridge Singapore 117542\\
 and \\
Mitch C. Warachka\\
School of Business,
Singapore Management University\\
Singapore 178899}
\date{}

\newtheorem{define}{Definition}[section]
\newtheorem{prop}{Proposition}[section]
\newcommand{\expon}{{\rm e}}

\begin{document}
\maketitle

\begin{abstract}
We investigate LIBOR-based derivatives using a parsimonious field
theory interest rate model capable of instilling imperfect
correlation between different maturities. Delta and Gamma hedge
parameters are derived for LIBOR Caps against fluctuations in
underlying forward rates. An empirical illustration of our
methodology is conducted to demonstrate the influence of
correlation on the hedging of interest rate risk.
\end{abstract}

\section{Introduction}

LIBOR-based derivatives such as Caps and Floors are important
financial contracts involving a sequence of quarterly payments ranging from one to
ten years. Consequently, pricing and hedging such derivatives requires the
modeling of multiple LIBOR rates.

In an economy where LIBOR rates are perfectly correlated across different
maturities, a single volatility function is sufficient. However,
non-parallel movements in the LIBOR term structure introduce an
important complication. To reduce the number of necessary inputs,
volatility parameters within certain time intervals are often
assumed to be identical. However, this assumption represents a
serious compromise, and longer maturity options still require a
large number of volatility parameters even after such aggregation.

In light of this issue, we utilize field theory models introduced
by Baaquie \cite{baaqbook} to instill imperfect correlation
between LIBOR maturities as a parsimonious alternative to the
existing theory. We derive the corresponding hedge parameters for
LIBOR Caplets for applications to risk management. We then
demonstrate the ease with which our formulation is implemented and
the implications of correlation on the hedge parameters.

Hedge parameters that minimize the risk associated with a finite
number of random fluctuations in the forward interest rates is
provided in Baaquie, Srikant, and Warachka \cite{baaqmar2}.
Previously, field theory research has focused on applications
involving traditional Heath, Jarrow, and Morton \cite{HJM} forward
interest rates, and on the pricing of LIBOR-based derivatives as is Baaquie
\cite{baalibor}. This paper extends the concept of stochastic
Delta hedging developed in Baaquie \cite{baaqbook} to the hedging of LIBOR
derivatives.

The remainder of this paper begins with a review of the field
theory model for pricing LIBOR derivatives. Section 3 then
investigates their corresponding hedge parameters, while Section 4 details their empirical
implementation. The conclusion follows in Section 5.

\section{Field Theory Model} \label{QF}

The introduction of imperfect correlation between all underlying
LIBOR rates is accomplished by the specification of a propagator for
interest rate dynamics. In terms of notation, $L(t,T)$
denotes the LIBOR rate at the current time $t$ between time $T$ and
$T+\ell$ in the future where $\ell=1/4$ year denotes the standard
3-month time interval between payoffs.

Since forward rates are the basis for LIBOR rates, we first
detail the Lagrangian underlying the evolution of forward rates.
Let $A(t,x)$ be a two dimensional field driving the
evolution of forward rates $f(t,x)$ through time
\begin{equation}
\frac{\partial f(t,x)}{\partial t}=\alpha(t,x)+\sigma(t,x)A(t,x)
\end{equation}
where $\sigma(t,x)$ and  $\alpha(t,x)$ denote their volatility and drift
velocity respectively.

Following Baaquie and Bouchaud \cite{baaqbou}, the Lagrangian of
the field is defined by three parameters, namely $\mu$ $\lambda$
and $\eta$.

\begin{define} \label{L} The Lagrangian which describes the evolution of instantaneous forward rates equals
\begin{equation}
\label{LA} {\cal L}[A] = -\frac{1}{2} \left\{A^2(t,z)+\frac{1}{\mu^2}
\left( \frac{\partial A(t,z)}{\partial z} \right)^2 +\frac{1}{\lambda^4}
  \left( \frac{\partial^{2}A(t,z)}{\partial^{2}z} \right)^2\right\} \, ,
\end{equation}
where psychological future time is defined by $z=(x-t)^{\eta}$.
\end{define}

The Lagrangian in Definition \ref{L} contains a squared Laplacian
term that describes the stiffness of the forward rate curve.
Baaquie and Bouchaud \cite{baaqbou} demonstrate that this
formulation is empirically able to account for the phenomenology
of interest rate dynamics. Ultimately, our pricing formulae for
Caps and Floors stems from a volatility function and correlation
parameters $\mu$, $\lambda$ and $\eta$ contained in the
propagator, as well as the initial term structure.

These forward rate dynamics are ultimately invoked for the pricing
of Caps and Floors after expressing derivatives on interest rates in
terms of their counterparts on bonds.

\subsection{LIBOR Dynamics} \label{Ldynamics}

The following is the relationship between the forward interest
rates and the LIBOR term structure
\begin{equation}
\label{libor}
L(t,T)=\frac{\expon^{\int^{T+\ell}_{T}dx f(t,x)}-1}{\ell} \, .
\end{equation}

In the original Heath, Jarrow, and Morton model \cite{HJM}, the martingale
measure is defined by discounting Treasury Bonds denoted $B(t,T)$ by
the money market account $R(t,t_*)$, defined as
\begin{equation}
R(t,t_*)=e^{\int_{t}^{t_*}r(t)dt} \, ,
\end{equation}
for the spot rate of interest denoted $r(t)$. In contrast, in this
paper all computations are carried out using the LIBOR measure for
which LIBOR rates evolve as martingales. In other words, for
$t_{*}>t$
\begin{eqnarray}
L(t,T_{n})=E_{L}\left[L(t_{*},T_{n})\right] \, .
\end{eqnarray}

Following the material in Baaquie \cite{baalibor}, the drift
$\alpha_{L}(t,x)$ that corresponds to the LIBOR martingale
condition is given by
\begin{equation}
\alpha_{L}(t,x)=-\sigma(t,x)\int_{T_{n}}^{x}dx^{'}D(x,x^{'};t)\sigma(t,x^{'})\,\quad;\quad
T_{n}\leq x<T_{n+\ell} \, .
\end{equation}

As proved in Baaquie \cite{baalibor}, a money market numeraire
entails more complex calculations but arrives at identical prices if
one instead uses the LIBOR measure. For the remainder of this paper,
the subscript of $L$ is suppressed with all expectations
performed under the LIBOR measure.

\subsection{Pricing an Individual Caplet} \label{pricing}

The existing literature justifies the Black model for pricing Caps
and Floors by modifying risk neutral Heath, Jarrow, and Morton
\cite{HJM} forward rates to yield LIBOR dynamics under the forward
measure. Brace, Gatarek, and Musiela \cite{BGM} is the seminal paper
in this area, with additional details found in Musiela and Rutkowski
\cite{MR}.

We review the field theory pricing formula for a Caplet for both a
general volatility function $\sigma(t,T)$ and propagator
$D(x,x^{'};t)$ underlying risk neutral forward rates. Denote the principal amount of the Cap as $V$. If
the Caplet is exercised at time $T$, the payment is made in arrears
at time $T+\ell$. Hence the payoff function at time $T+\ell$ is
given by
\begin{eqnarray}
g(T+\ell)=\ell V \left( L(T,T) -K \right)_{+} \,
\end{eqnarray}
where $K$ denotes the strike rate of the Caplet. Note that before
discounting the payoff at time $T$, we first discount from
$T+\ell$ back to time $T$. The entire expression for the Caplet
price is given by
\begin{eqnarray}
Caplet(t,T)&=&B(t,T)E_{[t,T]}\left[B(T,T+\ell)g(T+\ell)\right]\\ \nonumber \\
&=&\left[\frac{V}{X}\right]B(t,T)E_{[t,T]}\left[\left(X-e^{-\int_{T}^{T+\ell}dx
f(T,x)}\right)_{+}\right]
\end{eqnarray}
according to equation (\ref{libor}) and for $X\equiv
1\big{/}(1+\ell K)$. Observe that invoking the forward measure
involves multiplying by the bond $B(t,T)$ with the only random
forward rate term structure from $T$ to $T+\ell$. Then,
\begin{eqnarray}
Caplet(t,T) &=&
\int_{-\infty}^{+\infty}d{G}\Psi(G,T,T+\ell)(X-e^{-G})_{+}
\end{eqnarray}
where, as the derivation in Baaquie \cite{baaqbook},
$\Psi(G,T,T+\ell)$ equals
\begin{eqnarray}
\left[\frac{V}{X}\right]B(t,T)\sqrt{\frac{1}{2\pi{q}^{2}(T-t)}}
\exp\left\{-\frac{1}{2q^{2}(T-t)}\left(G-\int_{T}^{T+\ell}dx{f}(t,x)-\frac{q^{2}(T-t)}{2}\right)^{2}\right\}
\, .\label{capprice}
\end{eqnarray}
The above result leads to the next proposition for Caplet pricing.

\begin{prop} \label{cap_price} The price of a Caplet with strike K which matures at time T equals
\begin{eqnarray}
Caplet(t,T,T+\ell)=\left[\frac{V}{X}\right]B(t,T)\left[X{N}(d_{+})-FN(d_{-})\right]
\end{eqnarray}
for $X=\frac{1}{1+\ell{K}}$ , $B(t,T)=\frac{1}{1+\ell{L(t,t,T)}}$,
and the following definitions
\begin{eqnarray}
F&=&\frac{1}{1+\ell{L}(t,T)}\nonumber\\
d_{\pm}&=&\frac{1}{q\sqrt{T-t}}\left[\ln\left(\frac{F}{X}\right)\pm\frac{q^{2}(T-t)}{2}\right]\nonumber\\
q^{2}&=&\frac{1}{T-t}\int_{t}^{T}d\tilde{t}\int_{T}^{T+\ell}dx{d}x^{'}\sigma(\tilde{t},x)D(x,x^{'};\tilde{t})\sigma(\tilde{t},x^{'})
\, . \label{qcap}
\end{eqnarray}
\end{prop}

Observe that the propagator for forward rates are elements of the
Caplet price. The price of an at-the-money Caplet is then defined for
$X=F$, which yields $d_{\pm}=\pm\frac{q\sqrt{T-t}}{2}$, implying
an associated price of
\begin{eqnarray}
Caplet(t,T,T+\ell)&=&V{B}(t,T)\left[N(-d_{-})-N(-d_{+})\right]\\
&=&V{B}(t,T)\left[N\left(\frac{q\sqrt{T-t}}{2}\right)-N\left(-\frac{q\sqrt{T-t}}{2}\right)\right] \, .
\end{eqnarray}

\section{Hedging}

This section details the implications of our field theory model on
hedging LIBOR derivatives. The impact of
correlation 
is examined in the context of the residual variance and the Delta
hedge parameter for a portfolio. In particular, the more
practical Stochastic Delta hedging technique is given in Subsection \ref{deltahedge}.

A portfolio $\Pi(t)$ composed of a $Cap(t,t_{*},T)$\footnote{This
is a more general expression for a Cap referred to as the midcurve
Cap.} and $N$ LIBOR futures contracts, with the futures chosen to ensure fluctuations in
the value of the portfolio are minimized, is studied. This portfolio equals
\begin{eqnarray}
\Pi(t)=Cap(t,t_{*},T) +\sum_{i=1}^N n_{i}(t){\cal F}(t,T_{i}) \, ,
\label{Pi}
\end{eqnarray}
where $n_{i}(t)$ represents the hedge parameter for the $i^{th}$
futures contract included in the portfolio. The LIBOR futures and
Cap prices are denoted by
\begin{eqnarray}
{\mathcal{F}}(t,T_{i})&=&  V [1-\ell{L}(t,T_{i})]\\
Cap(t,t_{*},T)&=& \tilde{V}
B(t,T)\int^{+\infty}_{-\infty}\frac{d{G}}{\sqrt{2\pi{q_{*}}^2}}e^{-\frac{1}{2q_{*}^2}
\left( G-\int_{T}^{T+\ell}dx{f}(t,x)-\frac{q_{*}^2}{2}
\right)^2}(X-e^{-G})_{+} \,  \label{eqncap}
\end{eqnarray}
where for the midcurve Cap, we have
\begin{eqnarray}
q_{*}^{2}&=&\frac{1}{t_{*}-t}\int_{t}^{t_{*}}d\tilde{t}\int_{t_{*}}^{t_{*}+\ell}dx{d}x^{'}\sigma(\tilde{t},x)D(x,x^{'};\tilde{t})\sigma(\tilde{t},x^{'}) \, .
\end{eqnarray}
From equation (\ref{Pi}), we have
\begin{eqnarray}
\Pi(t)=Cap(t,t_{*},T)+V\sum_{i=1}^{N}n_{i}(t)(1-\ell L(t,T_{i}))
\, . \nonumber
\end{eqnarray}
For the sake of brevity, we suppress $V\sum_{i=1}^{N}n_{i}$ in the
above equation, which is irrelevant for hedging, and change the
negative sign before the LIBOR futures to positive as follows
\begin{eqnarray}
\Pi(t)&=&Cap(t,t_{*},T)+V\sum_{i=1}^{N}n_{i}(t)\ell L(t,T_{i}))\nonumber\\
&=&Cap(t,t_{*},T)+V\sum_{i=1}^{N} n_{i}(t) \left(
e^{\int_{T_{i}}^{T_{i}+\ell}f(t,x)}-1 \right) \, .\label{Pisimply}
\end{eqnarray}

\subsection{Residual Variance} \label{residueV}

Hedging a Cap denoted $Cap(t,T)$ using LIBOR futures contracts can
be accomplished by minimizing the residual variance of the hedged
portfolio. It is the instantaneous {\it change} in the portfolio
value that is stochastic. Therefore, the volatility of this change
is computed to ascertain the efficacy of the hedge portfolio.

The variance of the portfolio change,
$Var\left[\frac{d\Pi(t)}{dt}\right]$, equals
\begin{eqnarray}
&& Var\left[\frac{dCap(t,T)}{dt}\right]+Var\left[\sum_{i=1}^{N}n_{i}\frac{dL(t,T_{i})}{dt}\right]
\nonumber \\ \nonumber \\ && + \sum_{i=1}^{N}n_{i}Var\left[ \left<
\frac{dCap(t,T)}{dt},\frac{dL(t,T_{i})}{dt} \right> -
\left<\frac{dCap(t,T)}{dt} \right> \left< \frac{dL(t,T_{i})}{dt}
\right> \right] \, .
\end{eqnarray}

The detailed calculation for determining the hedge parameters and
portfolio variance is carried out in the Appendix. As in Baaquie,
Srikant, and Warachka \cite{baaqmar2}, the following notation is
introduced for simplicity
\begin{eqnarray}\label{not1}
K_i &=& \chi \hat{L}(t,T_i) \int_T^{T+\ell} dx
\int_{T_i}^{T_i+\ell}
dx' \sigma(t, x) \sigma(t, x') D(x,x';t) \, , \nonumber \\
M_{ij} &=& \hat{L}(t,T_i) \hat{L}(t,T_j) \int_{T_i}^{T_i+\ell} dx
\int_{T_j}^{T_j+\ell} dx' \sigma(t, x) \sigma(t, x') D(x,x';t) \, .
\end{eqnarray}
Equation (\ref{not1}) allows the residual variance in equation
(\ref{eq:hedgingcap}) to be succinctly expressed as
\begin{eqnarray}
 \label{eq:res}
&&\chi^2 \int_T^{T+\ell} dx \int_T^{T+\ell} dx' \sigma(t, x)
\sigma(t, x') D(x,x';t) + 2  \sum_{i=1}^N \Delta_i K_i +
\sum_{i=1}^N \sum_{j=1}^N \Delta_i \Delta_j M_{ij}
\end{eqnarray}
which contains covariance terms. When at-the-money, the value of $\chi$ below
facilitates our empirical estimation of the model in Section 4
\begin{eqnarray}
\chi&=& - V B(t,T)
\int^{+\infty}_{-\infty}\frac{d{G}}{\sqrt{2\pi{q}^2}}\frac{1}{q^2}
\left( G-\int_{T}^{T+\ell}dx{f}(t,x)-\frac{q^2}{2}
\right)\nonumber\\
&&\times\left\{e^{-\frac{1}{2q^2} \left( G-\int_{T}^{T+\ell}dx{f}(t,x) -\frac{q^2}{2} \right)^2}(X-e^{-G})_{+}\right\}\nonumber\\
&=&VB(t,T)\left\{\frac{1}{\sqrt{2\pi
q^2}}e^{-1/2d_{+}^2}+\left(\frac{1+\ell{K}}{1+\ell{L}}\right)\left[-\frac{1}{\sqrt{2\pi{q}^2}}e^{-1/2d_{-}^2}+N(d_{-})\right]\right\}
\,
\end{eqnarray}
where  $d_{\pm}=\left(\ln\frac{X}{F}\pm{q}^{2}/2\right)/q$. The
value of $\chi$ for an at-the-money options yields $d_{\pm}=\pm q/2$ which implies
\begin{equation}
\chi(t,T)|_{ \text{at-the-money}} = VB(t,T)N(d_{-}) \, .
\end{equation}

Observe that the residual variance depends on the correlation
between forward rates described by the propagator. Ultimately, the
effectiveness of the hedge portfolio is an empirical question
since perfect hedging is not possible without shorting the
original bond. This empirical question is addressed in Section
\ref{Empirical} when the propagator is calibrated to market data.

Hedge parameters $n_i$ that minimize the residual variance in
equation (\ref{eq:res}) are
\begin{eqnarray}\label{dg}
n_i &=& - \sum_{j=1}^N K_j M_{ij}^{-1} \, .
\end{eqnarray}
These parameters represent the optimal amounts of the futures contracts to include
in the hedge portfolio.

Equation (\ref{dg}) is proved by differentiating equation
(\ref{eq:res}) with respect to $n_i$ and subsequently solving for
its value. The variance of the hedged portfolio in equation (\ref{rv}) is
proved by substituting the result of equation (\ref{dg}) into equation
(\ref{eq:res})
\begin{eqnarray}\label{rv}
  V_{R} &=& \chi^2 \int_T^{T+\ell} dx \int_T^{T+\ell} dx' \sigma(t, x) \sigma(t,x') D(x,x';t) - \sum_{i=1}^N \sum_{j=1}^N K_i M_{ij}^{-1} K_j
\end{eqnarray}
which declines monotonically as $N$ increases.

The residual variance in equation (\ref{rv}) enables the
effectiveness of the hedge portfolio to be evaluated. Therefore,
equation (\ref{rv}) is the basis for studying the impact of
including different LIBOR futures contracts in the hedge portfolio. For $N=1$, a single
maturity $T_i$ is evaluated, and the residual variance in equation
(\ref{rv}) reduces to
\begin{eqnarray}
&& \chi^2  \int_T^{T+\ell} dx \int_T^{T+\ell} dx' \sigma(t, x) \sigma(t,x') D(x,x';t) \nonumber \\
&-& \left( \frac{ \left(\int_T^{T+\ell} dx \int_{T_1}^{T_1+\ell}
dx' \sigma(t, x) \sigma(t, x') D(x,x';t) \right)^2}{
\int_{T_1}^{T_1+\ell} dx \int_{T_1}^{T_1+\ell} dx' \sigma(t, x)
\sigma(t, x') D(x,x';t)} \right)  \, . \label{eq:hed}
\end{eqnarray}
The second term in equation (\ref{eq:hed}) represents the
reduction in variance attributable to the hedge portfolio. To
obtain the HJM limit, the propagator is constrained to equal one,
reducing the residual variance $V_R$ in equation (\ref{eq:hed})
\begin{eqnarray}\label{HJM}
\chi^2 \left[ \left( \int_T^{T+\ell} dx \sigma(t, x) \right)^{2} -
\left( \frac{ (\int_T^{T+\ell} dx \int_{T_1}^{T_1+\ell} dx'
\sigma(t, x) \sigma(t, x'))^2}{\int_{T_1}^{T_1+\ell} dx
\sigma(t,x) \sigma(t, x')} \right) \right] \,
\end{eqnarray}
to zero. This HJM limit is consistent with our intuition
that the residual variance is identical zero for any LIBOR
maturity  since all forward rates are perfectly correlated. This
result is also shown empirically in Section \ref{Empirical}.
However, results from hedging with two LIBOR futures contracts in
HJM model are not presented since one degree of freedom cannot be
hedged with two instruments. Indeed, in this circumstance,
$M_{ij}^{-1}$ is singular.

\subsection{Stochastic Hedging} \label{deltahedge}

Residual variance enables us to control the effectiveness of the
hedging procedure. However, instead of only hedging downward movements in the Cap price, residual variance operates on
all forward rate fluctuations, including
those that increase the portfolio's value. For this reason, we
study stochastic hedging which is more practical since we can decide which forward rates
to hedge against.

Stochastic hedging of interest rate derivatives has been
introduced by Baaquie \cite{baaqbook}, where the specific case of
hedging Treasury Bonds is considered in detail. We focus on
applying this technique to the hedging of a LIBOR Cap. Consider
the hedging of a Cap against fluctuations in the forward rate
$f(t,x)$. A portfolio $\Pi(t)$ composed of a $Cap(t_{0},t_{*},T)$
and one LIBOR futures contract is studied.

As in equation (\ref{Pisimply}), we set $N=1$ to obtain
\begin{eqnarray}
\Pi(t) =Cap(t,t_{*},T)+V n_{1}(t) \left(
e^{\int_{T_{1}}^{T_{1}+\ell}f(t,x)}-1 \right) \, . \nonumber
\end{eqnarray}
The portfolio is required to be
independent of small changes in the forward rate. Thus, Delta
hedging this portfolio requires
\begin{eqnarray}
\frac{\delta}{\delta{f(t,x)}}\Pi(t)=0 \, . \label{delta}
\end{eqnarray}
In field theory, for each time $t$, there are infinitely many
random variables driving forward rates, and one can never exactly
Delta hedge by satisfying equation (\ref{delta}). The best
alternative is to Delta hedge on average, and this scheme is
referred to as stochastic Delta hedging as detailed in Baaquie \cite{baaqbook}.
To implement stochastic Delta hedging, one considers the
conditional expectation value of the portfolio $\Pi(t)$,
conditioned on the occurrence of some specific value of the
forward rate $f_{h}\equiv f(t,x_{h})$, namely $E[\Pi(t)|f(t,x_{h})]$.
Define the conditional probability of a Cap and a LIBOR futures by
\begin{eqnarray}
\tilde{Cap}
(t,t_{*},T;f_{h})&=&E[Cap(t,t_{*},T)|f_{h}] \\
\nonumber \tilde{L}(t,T_{1};f_{h})&=&E[L(t,T_{1})|f_{h}] \, . \,
\end{eqnarray}

From Baaquie \cite{baaqbook} and equation (\ref{capprice}), we have
the conditional probability of a Cap given by
\begin{eqnarray}\label{condcap}
\tilde{Cap}(t,t_{*},T;f_{h})&=&\tilde{V}\int_{-\infty}^{\infty}dG\left\{(x-e^{G})_{+}\Psi(G|f_{h})\right\}\label{capcon}   \\
\Psi(G|f_{h})&=&\frac{\int_{-\infty}^{\infty}\frac{dp}{2\pi}
e^{-\frac{q_{h}^{2}}{2}p^{2}}e^{ip(G-\frac{q_{h}^{2}}{2})}\int{D}f
e^{-\int_{t}^{T}f(t,x)}e^{ip\int_{T}^{T+l}dx
f(t,x)}\delta(f({t,x_{h}})-f)e^{S}}{\int{D}f\delta(f({t,x_{h}})-f)e^{S}}
\, ,\nonumber
\end{eqnarray}
while the conditional probability of a LIBOR futures is
\begin{eqnarray}\label{condlibor}
\tilde{L}(t,T_{1};f_{h})&=&\int_{-\infty}^{\infty}dG e^{G}\Phi(G|f;t,T_{1})\label{liborcon}  \nonumber \\
\Phi(G|f;t,T_{1})&=&\frac{\int{D}f\delta(G-\int_{T_{1}}^{T_{1}+\ell}f(t,x)dx)\delta(f(t,x_{h})-f)e^{S}}
{\int{D}f\delta(f(t,x_{h})-f)e^{S}} \, .
\end{eqnarray}

Stochastic Delta hedging is defined by approximating equation
(\ref{delta}) as
\begin{eqnarray}
\frac{\partial}{\partial{f_{h}}}E[\Pi(t)|f_{h}] &=& 0 \,
.\label{stodelta}
\end{eqnarray}
Hence, from equation (\ref{stodelta}), stochastic Delta hedging yields
\begin{eqnarray}
n_{1} &=& -\frac{\partial{\tilde{Cap}(t,t_{*},T;f_{h})}}
{\partial{f_{h}}}/\frac{\partial{\tilde{L}(t,T_{1};f_{h})}}{\partial{f_{h}}}
\, . \label{parameter}
\end{eqnarray}
Thus, changes in the hedged portfolio
$\Pi(t)$ are, on average, sensitive to fluctuations in the forward rate
$f(t,x_{h})$.

The conditional probability in equation (\ref{condcap}) and equation
(\ref{condlibor}) along with the hedge parameter $n_{1}$ is evaluated explicitly
for the field theory description of forward rates in the Appendix which also contains
the relevant notation. One should notice that nontrivial
correlations appear in all the terms. The final result, from
equation (\ref{hedgweight}), is given by
\begin{eqnarray}
n_{1}=\frac{C\cdot{\tilde{Cap}}(t,t_{*},T;f_{h})-B\cdot\chi\cdot\tilde{V}\cdot
\left[
XN^{'}(d_{+})/Q+e^{-G_{0}+\frac{Q^2}{2}}N(d_{-})-e^{-G_{0}+\frac{Q^2}{2}}N^{'}(d_{-})/Q
\right] }{e^{G_{1}+\frac{Q_{1}^2}{2}}\cdot{B}_{1}} \, .
\end{eqnarray}
As a comparison, the HJM limit is also
analyzed in the Appendix.

Furthermore, one can Gamma hedge the same forward rate. To
hedge against the $\partial^{2}\Pi(t)/\partial f^2$
fluctuations, one needs to form a portfolio with two LIBOR futures
contracts that minimizes the change in the value of
$E[\Pi(t)|f_{h}]$ by both Delta and Gamma hedging. These
parameters are solved analytically, with empirical results
presented in Section 4.

Suppose a Cap needs to be hedged against the fluctuations of two
forward rates, namely $f(t,x_{i})$ for $i=1,2$. The conditional
probabilities for the Cap and LIBOR futures, with two forward
rates fixed at $f(t,x_{i})=f_{i}$, are
\begin{eqnarray}
\tilde{Cap}(t,t_{*},T;f_{1},f_{2})&=&E[Cap(t,t_{*},T)|f_{1},f_{2}]\nonumber\\
\tilde{L}(t,T_{1};f_{1},f_{2})&=&E[L(t,T_{1})|f_{1},f_{2}] \, .
\nonumber
\end{eqnarray}
A portfolio of two LIBOR futures contracts with different
maturities $T_{i}\neq T$ is defined as
\begin{eqnarray}\label{hedge2forward}
\Pi(t)=Cap(t,t_{*},T)+\sum_{i=1}^{N}n_{i}(t)L(t,T_{i}) \, ,
\end{eqnarray}
where the hedging of this portfolio at instant time $t$ is given by
\begin{eqnarray}
\delta\Pi(t,f_{1},f_{2})=\frac{\partial\Pi}{\partial t}\delta
t+\sum_{i=1}^{2}\frac{\partial\Pi}{\partial f_{i}}\delta
f_{i}+\frac{1}{2}\sum_{i=1}^{2}\frac{\partial^{2}\Pi}{\partial
f_{i}^{2}}\delta^{2}f_{i}+\frac{1}{2}\frac{\partial^{2}\Pi}{\partial
f_{1}\partial f_{2}}\delta f_{1}\delta
f_{2}+O(\epsilon^{2})\label{hedging}
\end{eqnarray}
with $\delta t\equiv\epsilon=1/360$ year, while higher orders of
$\epsilon$ are negligible. Furthermore, the dynamics
$\dot{\Pi}=\delta\Pi/\delta t$ equal
\begin{eqnarray}
\dot{\Pi}(t,f_{1},f_{2})=\frac{\partial\Pi}{\partial
t}+\sum_{i=1}^{2}\frac{\partial\Pi}{\partial f_{i}}\dot{
f}_{i}+\frac{\epsilon}{2}\sum_{i=1}^{2}\frac{\partial^{2}\Pi}{\partial
f_{i}^{2}}\dot{f}^{2}_{i}+\frac{\epsilon}{2}\frac{\partial^{2}\Pi}{\partial
f_{1}\partial f_{2}}\dot f_{1}\dot f_{2}+O(\epsilon) \, .
\end{eqnarray}
Since $\langle\dot{f}\dot{f}\rangle\sim \frac{1}{\epsilon}$ as in
Baaquie \cite{baaqbook}, $\epsilon\dot{f}_{i}^{2}\sim
0(1)\sim\epsilon\dot{f}_{1}\dot{f}_{2}$, the second order terms
are as important as the first order terms.
Normal calculus retains the first order terms since
$\epsilon$ is infinitesimally small. However, $\epsilon=1$ day in our context.

The stochastic Delta hedging conditions are given by
\begin{eqnarray}
\frac{\partial}{\partial{f_{j}}}E[\Pi(t)|f_{1},f_{2}]=0\, \mbox{
for }  \, j=1,2  \nonumber
\end{eqnarray}
while stochastic Gamma hedging involves
\begin{eqnarray}
\frac{\partial^{2}}{\partial{f_{j}^{2}}}E[\Pi(t)|f_{1},f_{2}]=0\,
\mbox{ for }  \, j=1,2 \nonumber
\end{eqnarray}
with \textbf{Cross Gamma} hedging
\begin{eqnarray}
\frac{\partial^{2}}{\partial{f_{1}}\partial{f_{2}}}E[\Pi(t)|f_{1},f_{2}]=0\,
\nonumber
\end{eqnarray}
being unique to this paper. This Cross Gamma hedging only
make sense in field theory models where movements in any
specific forward rate can be hedged.

One can solve the above system of $N$ simultaneous equations to
determine the $N$ hedge parameters denoted $n_{i}$. The volatility
of the hedged portfolio is reduced by increasing the number of
forward interest rates being hedged.

For this portfolio, we can analytically prove that Delta hedge
parameters for the two forward rates differ by a prefactor
\begin{eqnarray}\label{predelta}
\frac{\partial}{\partial{f_{1}}}E[\Pi(t)|f_{1},f_{2}]=-\frac{A_{2}}{A_{12}}\frac{\partial}{\partial{f_{2}}}E[\Pi(t)|f_{1},f_{2}]=0
\end{eqnarray}
where $A_{2}$ and $A_{12}$ are defined in Appendix D. Therefore, Delta hedging against two forward rates can only determine the
portfolio including one LIBOR futures. Furthermore, Gamma hedging
two forward rates is the same except for a prefactor.

Overall, for hedging against two forward rates we are left with
three independent constraints from the above six constraints. In
order to study the effect of each set of constraints separately,
we form portfolios which include two LIBOR futures, and adopt
hedging strategies that involve more than Delta hedging to fix the
two hedge parameters. The first strategy implements one Delta
hedge and one Gamma hedge on one forward rate. The two hedging
parameters can also be fixed by one stochastic Delta hedge and an
additional Cross Gamma hedge.

All of these hedge strategies are evaluated explicitly in the
Appendix. Intuitively, we expect the portfolio to be hedged more
effectively with the inclusion of the Cross Gamma parameter.
Generally speaking, the field theory framework allows us to form portfolios
that include more LIBOR futures and hedge against any number of
forward rates.

Until now, we obtained the parameter for each choice of the LIBOR
futures and forward rates being hedged. Furthermore, we can
minimize the following
\begin{eqnarray}
\sum_{i=1}^{N} \left| n_{i} \right| \,   \label{portfoliozero}
\end{eqnarray}
to find the {\it minimum} portfolio. This additional constraint finds the most
effective futures contracts, where effectiveness is measured by requiring the smallest amount of contracts.

In general, stochastic Delta hedging against $N$ forward rates for
large $N$ is complicated, and closed-form solutions are difficult to
obtain.

\section{Empirical Implementation} \label{Empirical}

This section illustrates the implementation of our field theory
model and provides preliminary results for the impact of
correlation on the hedge parameters. The correlation parameter for
the propagator of LIBOR rates is estimated from historical data
on LIBOR futures and at-the-money options. We calibrate the term
structure of the volatility, $\sigma(\theta)$, (see \cite{data},
\cite{data1}) and the propagator with the parameters $\lambda$ and
$\mu$ as in Baaquie and Bouchaud \cite{baaqbou}. All the empirical
results showed below are calculated from the derivation expressed
in this paper.

\subsection{Empirical Results on Residual Variance}

\begin{figure}[h]
  \centering
  \epsfig{file=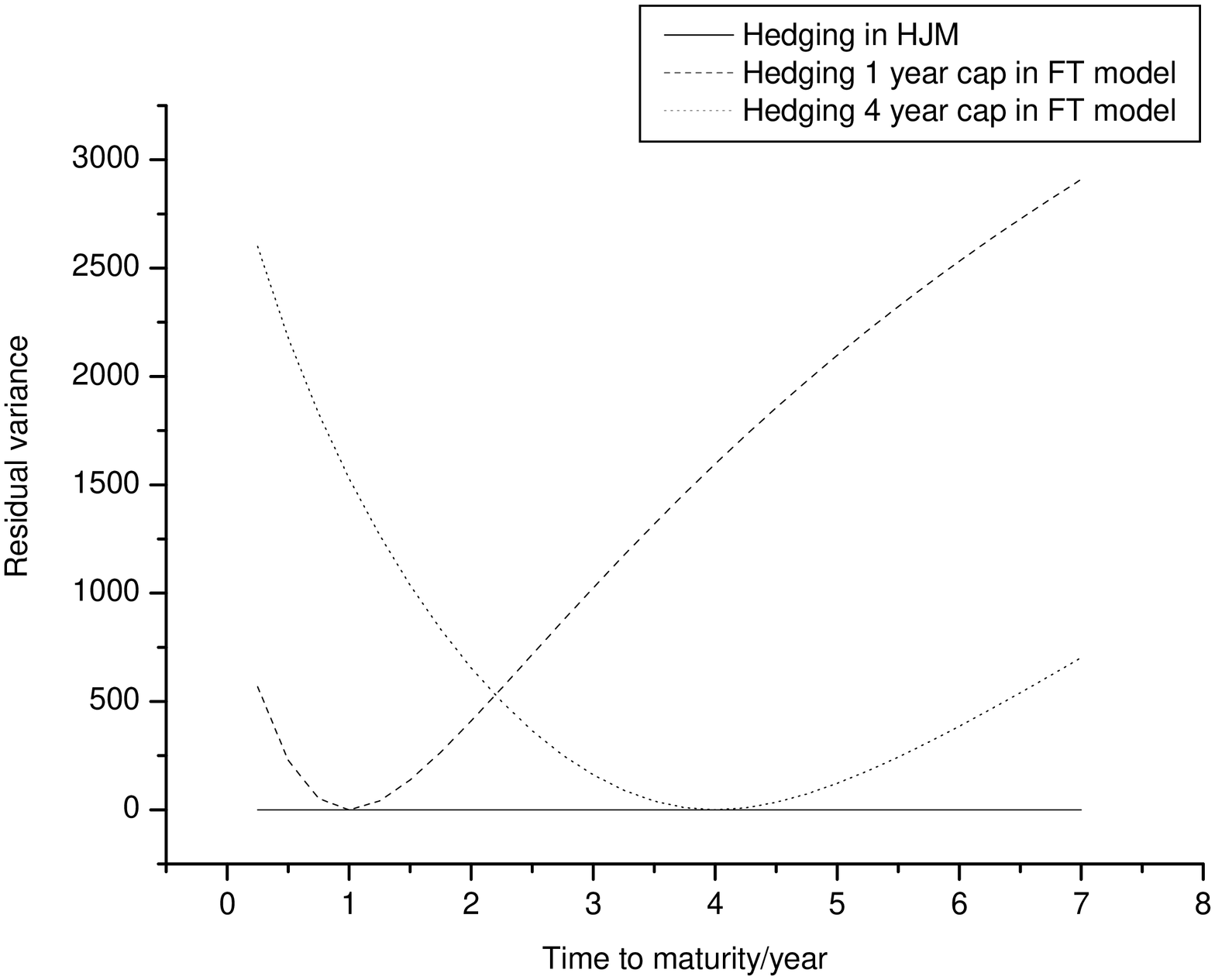, height=12cm, angle=-90}
  \caption{\emph{Residual variance $Var\left[\frac{d\Pi(t)}{dt}\right]$ for a one and four year Cap versus LIBOR futures
  maturity $T_{1}$ used to hedge portfolio} $\Pi(t)=Cap(t,t_{*}) + n_{1}(t){\cal F}(t,T_{1})$.}
  \label{fig:RV 1 by 1}
\end{figure}

The reduction in variance achievable by hedging a Cap with LIBOR
futures is the focus of this section. The portfolio
\begin{eqnarray}
\Pi(t)=Cap(t,t_{*},T) +\sum_{i=1}^N n_{i}(t){\cal F}(t,T_{i}) \,
\nonumber
\end{eqnarray}
is considered with $Var\left[\frac{d\Pi(t)}{dt}\right]$ being
minimized. The residual variance for hedging a 1 and 4 year Cap
with a LIBOR futures is shown in Fig. \ref{fig:RV 1 by 1}, along
with its HJM counterpart. Observe that the residual variance drops
to exactly zero when the same maturity LIBOR futures is used to
hedge the Cap.
\begin{figure}[h]
  \centering
  \epsfig{file=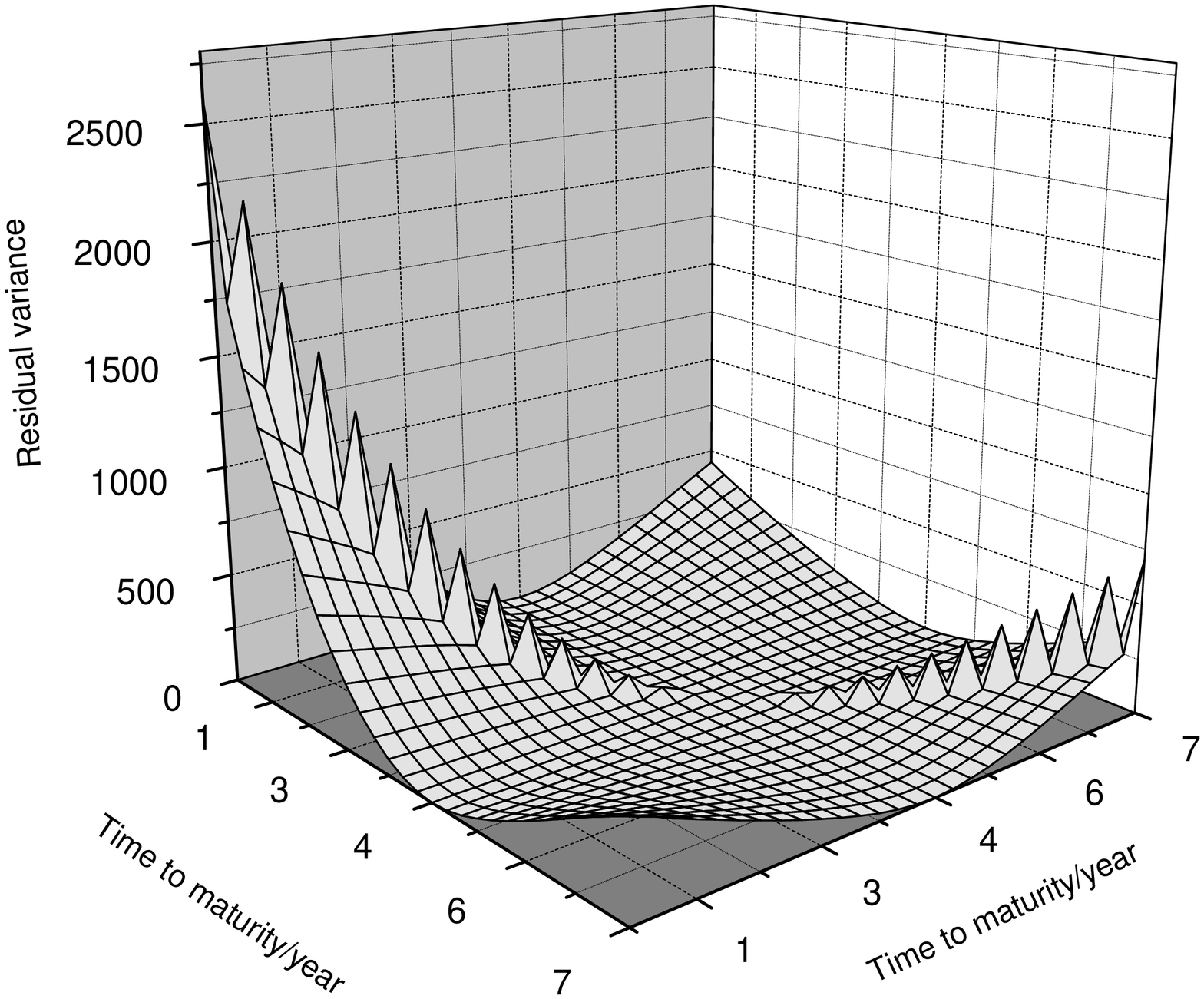, height=12cm, angle=-90}
  \caption{\emph{Residual variance $Var\left[\frac{d\Pi(t)}{dt}\right]$ for a four year Cap versus two
  LIBOR futures maturities $T_{i}$ used to hedge  $\Pi(t)=Cap(t,4) + \sum_{i=1}^{2}n_{i}(t){\cal F}(t,T_{i})$.}}
  \label{fig:RV 4 by 2}
\end{figure}

By considering the changes of residual variance with respect of
parameters $\lambda$ and $\mu$, we find the neighboring points
create no disparities, at least one cannot tell which offers the
better hedge. An explanation of this is effect is that forward
rates with similar maturities are strongly correlated.
Furthermore, the HJM residual variance for both hedging a 1 year
and 4 year Cap are identical to the residual variance=0 axis. This
is consistent with our analytical result in equation (\ref{HJM}).

The residual variance for hedging a 4 year Cap with two LIBOR
futures is provided in Fig. \ref{fig:RV 4 by 2}.
It is interesting to note that hedging with two instruments, even
with similar maturities, entails a significant decrease in
residual variance compared to hedging with one futures. This
is illustrated in Fig. \ref{fig:RV 4 by 2} where
$\theta=\theta^{'}$ represents hedging with one LIBOR futures. The
residual variance in this situation is higher than the nearby
points, and increases  in a discontinuous manner.

\subsection{Empirical Results on Stochastic Hedging}

\begin{figure}[h]
  \centering
  \epsfig{file=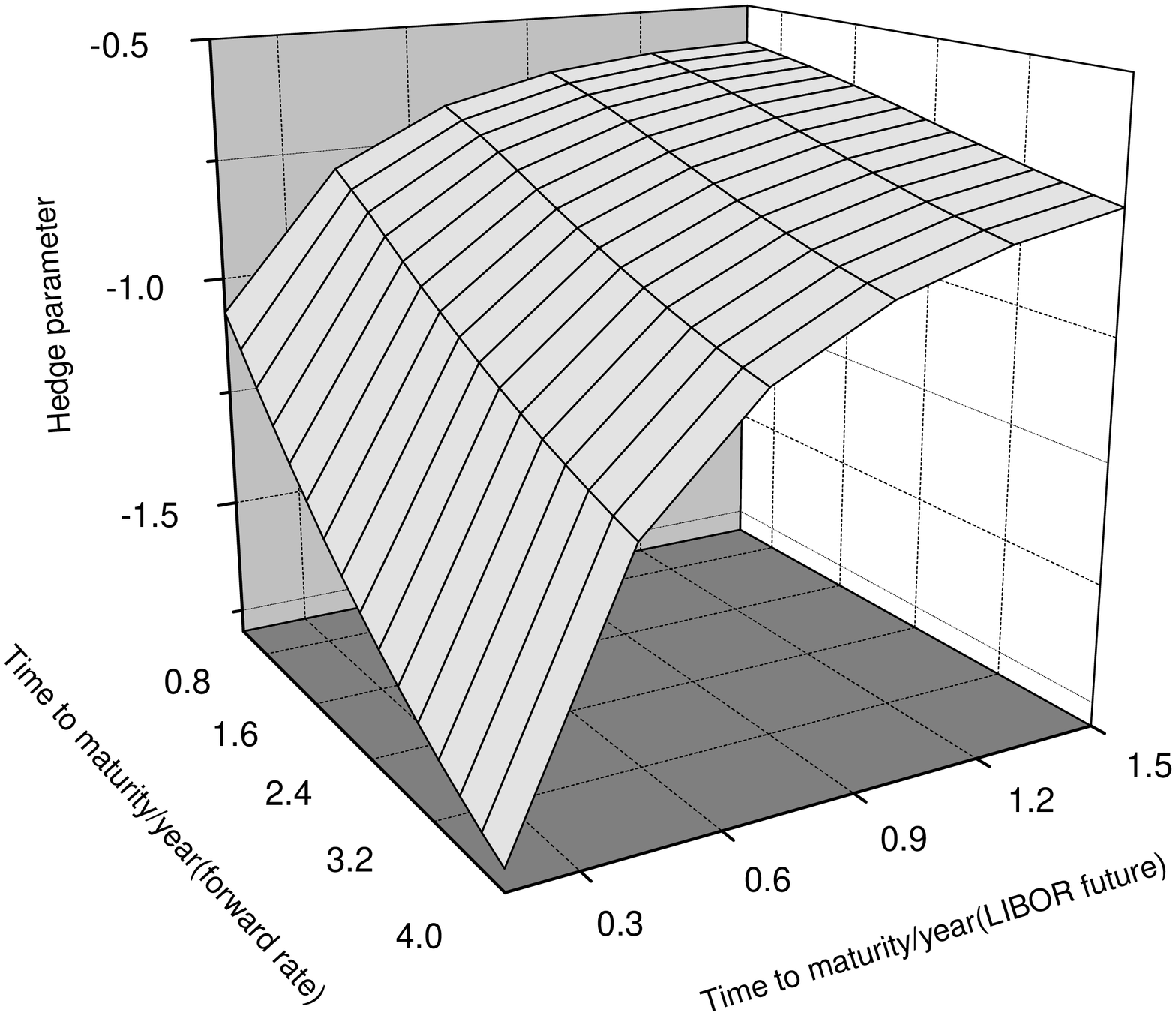, height=12cm ,angle=-90}
  \caption{\emph{Hedge parameter $n_{1}$ for stochastic Delta hedging of $Cap(t,1,4)$ using LIBOR futures
  maturity $T_{1}$ and forward rate maturity $x_{h}$ involving $\Pi(t)=Cap(t,1,4) + n_{1}(t){\cal F}(t,T_{1})$.}}
  \label{fig:stohedge}
\end{figure}
\begin{figure}[h]
  \centering
  \epsfig{file=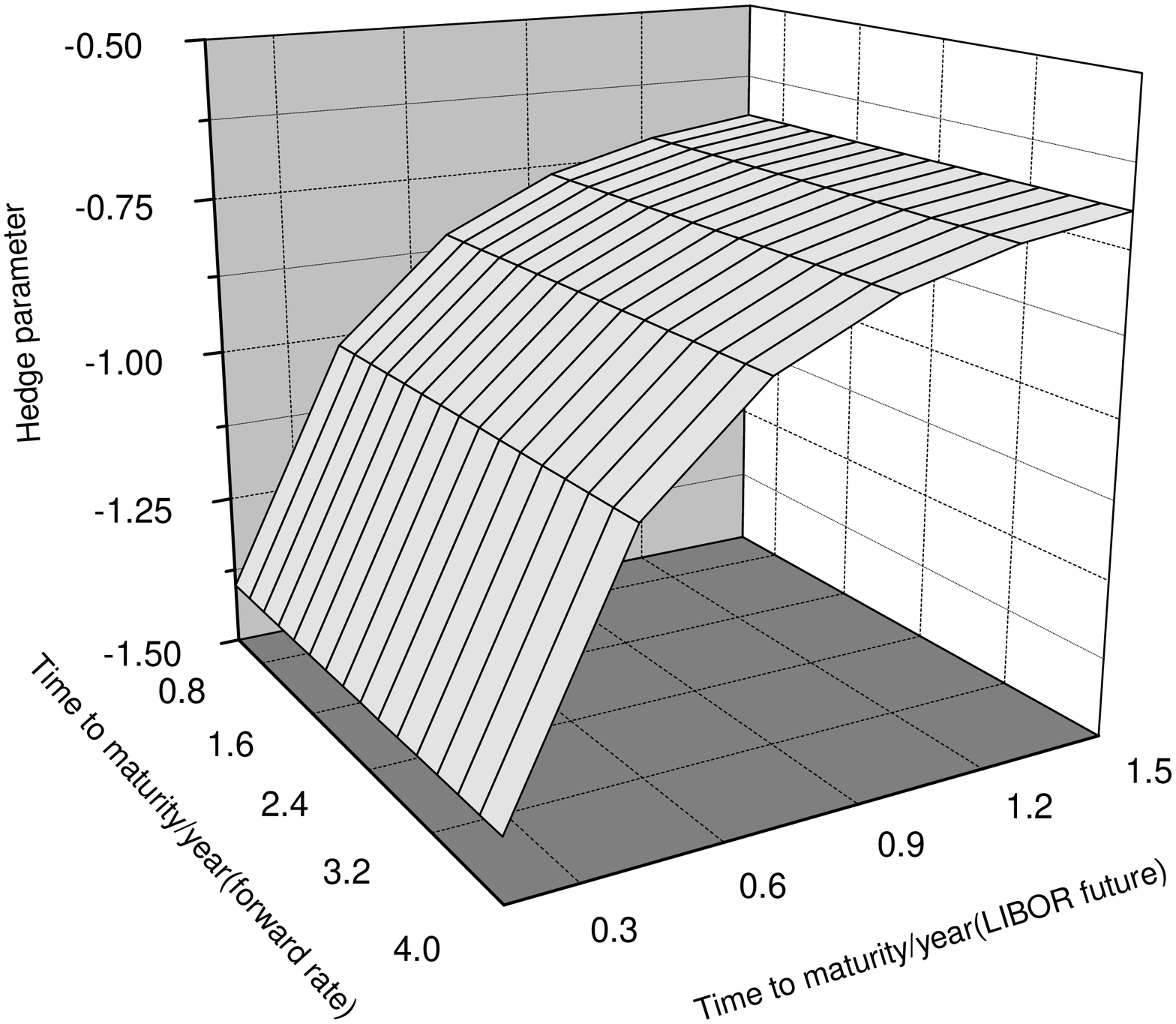, height=12cm ,angle=-90}
  \caption{\emph{Hedge parameter $n_{1}$ for stochastic hedging of $Cap(t,1,4)$ using LIBOR futures maturity $T_{1}$ and forward rate maturity $x_{h}$  in the HJM limit of $D=1$ (forward rates perfectly correlated) involving $\Pi(t)=Cap(t,1,4) + n_{1}(t){\cal F}(t,T_{1})$.}}
  \label{fig:stohedgeHJM}
\end{figure}

Stochastic hedging mitigates the risk of fluctuations in specified
forward rates. The focus of this section is on the stochastic
hedge parameters $n_{i}$, with the best strategy chosen to ensure
the LIBOR futures portfolio involves the smallest possible long
and short positions since $\sum_{i=1}^{N}|n_{i}|$ is minimized.

\subsubsection{Hedging in Field Theory Models Compared to HJM}

The comparison is carried out in the simplest portfolio where one
forward rate is hedged by one LIBOR futures, with a detailed
empirical study in Subsection \ref{1f1l}.  As an illustration,
Fig. \ref{fig:stohedge} plots the hedge parameter $n_{1}$ in our
field theory model against the LIBOR futures maturity $T_{1}$, and
the forward rate maturity $x_{h}$ being hedged. One advantage of
the field theory model is that, in principle, a hedge strategy
against the movements of infinitely many correlated forward rates
is available. To illustrate the contrast between our field theory
model and a single-factor HJM model, we plot the identical hedge
portfolio as above when $D=1$, which has been shown to be the HJM
limit of field theory models. From Fig. \ref{fig:stohedgeHJM}, for
the HJM limit, the hedge parameter $n_{1}$ is invariant to the
forward rate maturity $x_{h}$, which is expected since all forward
rates $f(t,x_{h})$ are perfectly correlated in a single-factor HJM
model. Therefore, it makes no difference which of the forward
rates is being hedged.

\subsubsection{Hedging Against One Forward Rate with One LIBOR
Futures}\label{1f1l}

We first study a portfolio with one LIBOR futures and one Cap to
hedge against a single term structure movement. The portfolio is
given by
\begin{eqnarray}
\Pi(t)=Cap(t,t_{*},T) +n_{1}(t){\cal F}(t,T_{1}) \, \nonumber
\end{eqnarray}
where the hedging is done by stochastic Delta hedging
$\frac{\partial}{\partial{f_{h}}}E[\Pi(t)|f_{h}]=0$ on forward
rate $f(t,x_{h})$.

Hedge parameters $n_{1}$ for different LIBOR futures maturities
$T_{1}$, and the forward rate maturity $x_{h}$, are shown in Fig.
\ref{fig:stohedge}. This figure describes the selection of the
LIBOR futures in the minimum portfolio that requires the fewest
number of long and short positions.

\begin{figure}[h]
  \centering
  \epsfig{file=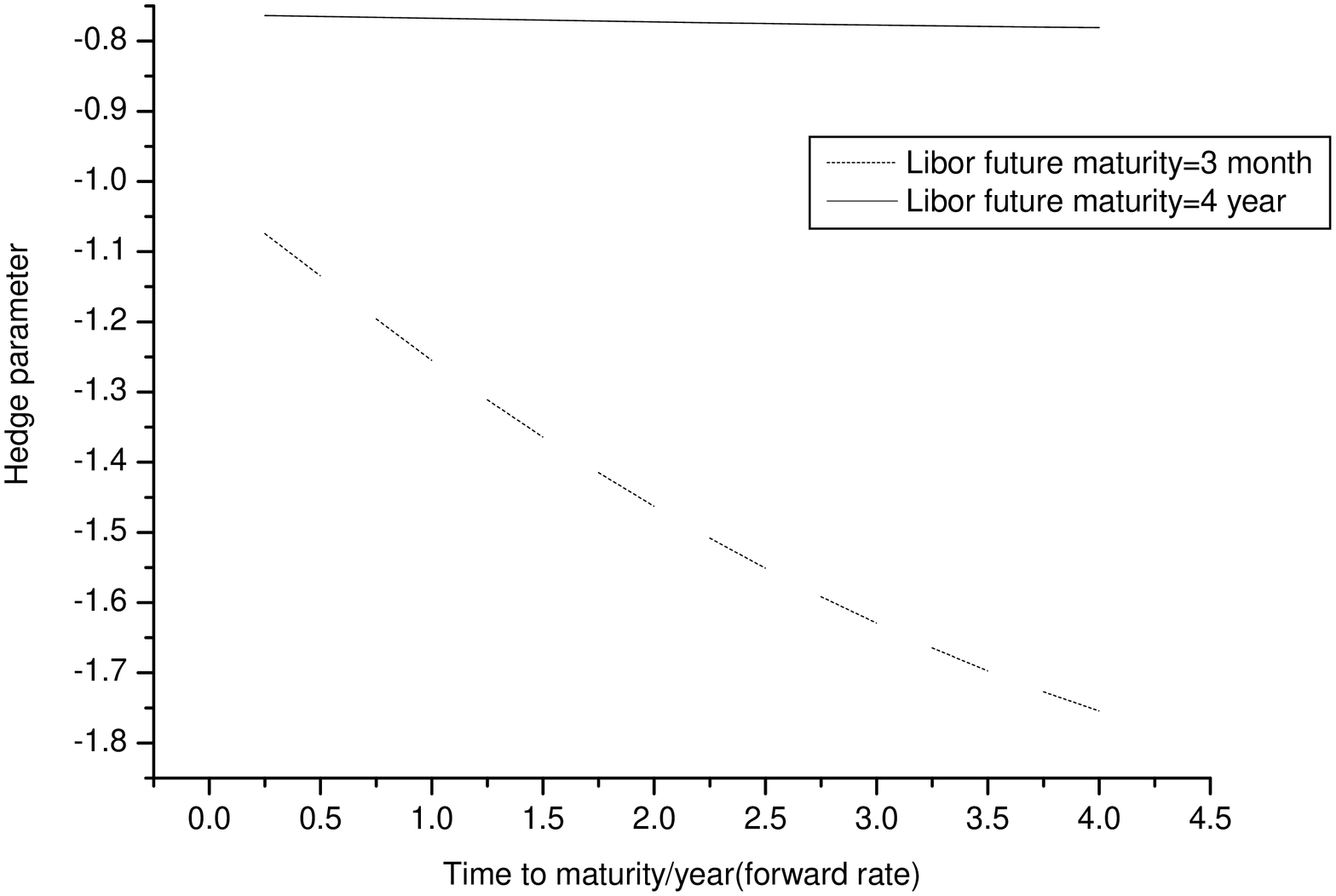, height=12cm ,angle=-90}
  \caption{\emph{Hedge parameter $n_{1}$ for stochastic hedging of $Cap(t,1,4)$ for forward
  maturity $x_{h}$ of forward rate $f(t,x_{h})$, with fixed LIBOR futures contract maturity $T_{1}$, involving  $\Pi(t)=Cap(t,1,4) + n_{1}(t){\cal F}(t,T_{1})$.}}
  \label{fig:stohedge2}
\end{figure}

\begin{figure}[h]
  \centering
  \epsfig{file=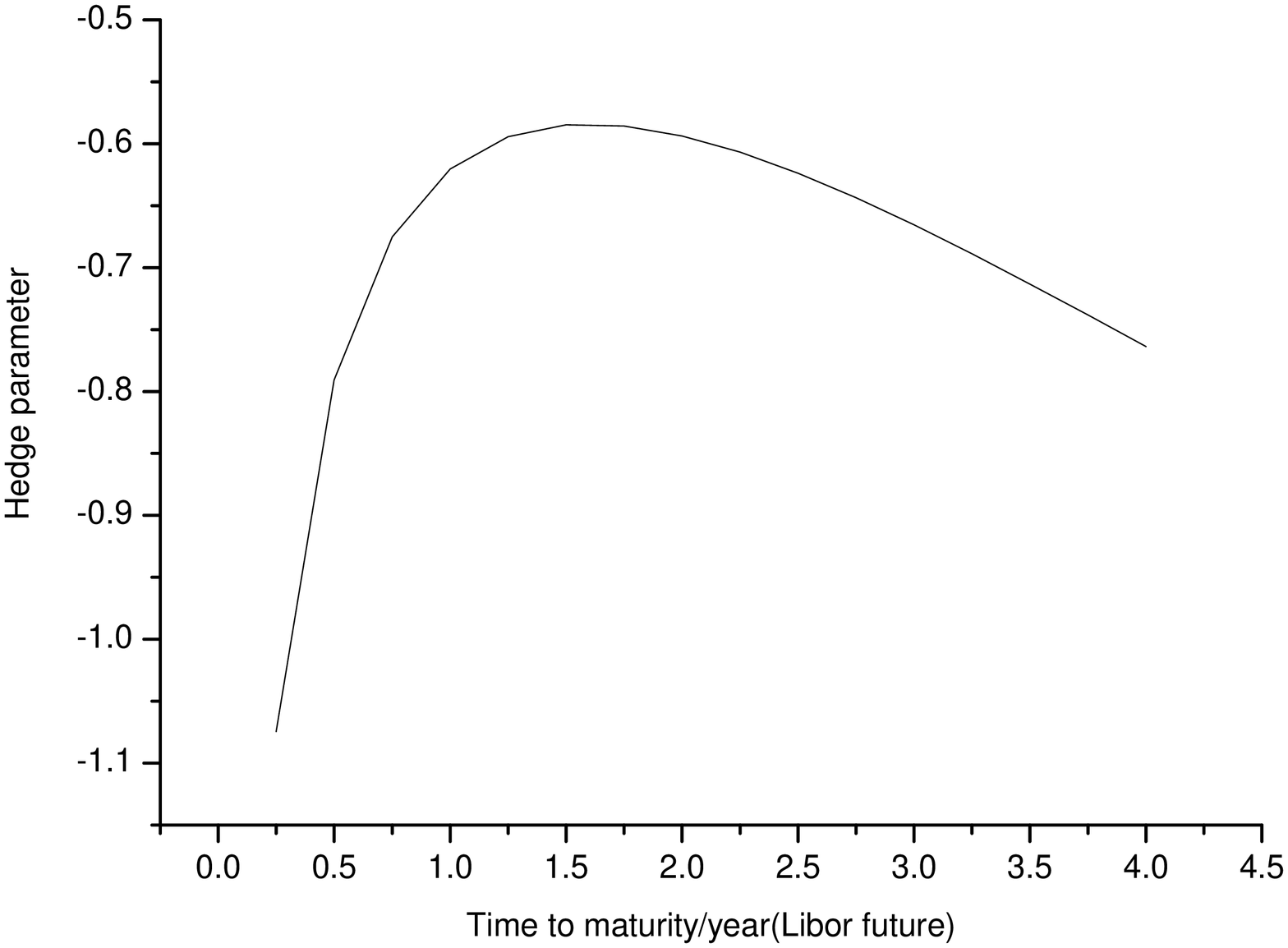, height=12cm ,angle=-90}
  \caption{\emph{Hedge parameter $n_{1}$ for stochastic hedging of $Cap(t,1,4)$ for LIBOR futures maturity $T_{1}$ when hedging
  against $f(t,t+\delta)$ with $\delta=3/12$, involving $\Pi(t)=Cap(t,1,4) + n_{1}(t){\cal F}(t,T_{1})$.}}
  \label{fig:stohedge1}
\end{figure}

Fig. \ref{fig:stohedge2} shows how the hedge parameters depend on
$x_{h}$ for a fixed $T$. Two limits $T_{1}= \delta  = \frac{1}{4}$
(3 months) and $T_{1}= 16 \delta$ are chosen. We find that $x_{h}=
\delta$ is always the most important forward rate to hedge
against. Another graph describing the parameter dependence on
$T_{1}$ is given in Fig. \ref{fig:stohedge1} with $x_h = \delta$.
The minimum of hedge parameter $n_{1}$ at $x_{h}\simeq 1.5$years
reflects the maximum of $\sigma(t,x)$ around the same future time.
For greater generality, we also hedge $Cap(t,t_{*},T)$ for
different $t_{*}$ and $T$ values, and find that although the value
of the parameter changes slightly, the shape of the parameter
surface is almost identical.

\subsubsection{Hedging Against One Forward Rate with Two LIBOR
Futures}

\begin{figure}[h]
  \centering
  \epsfig{file=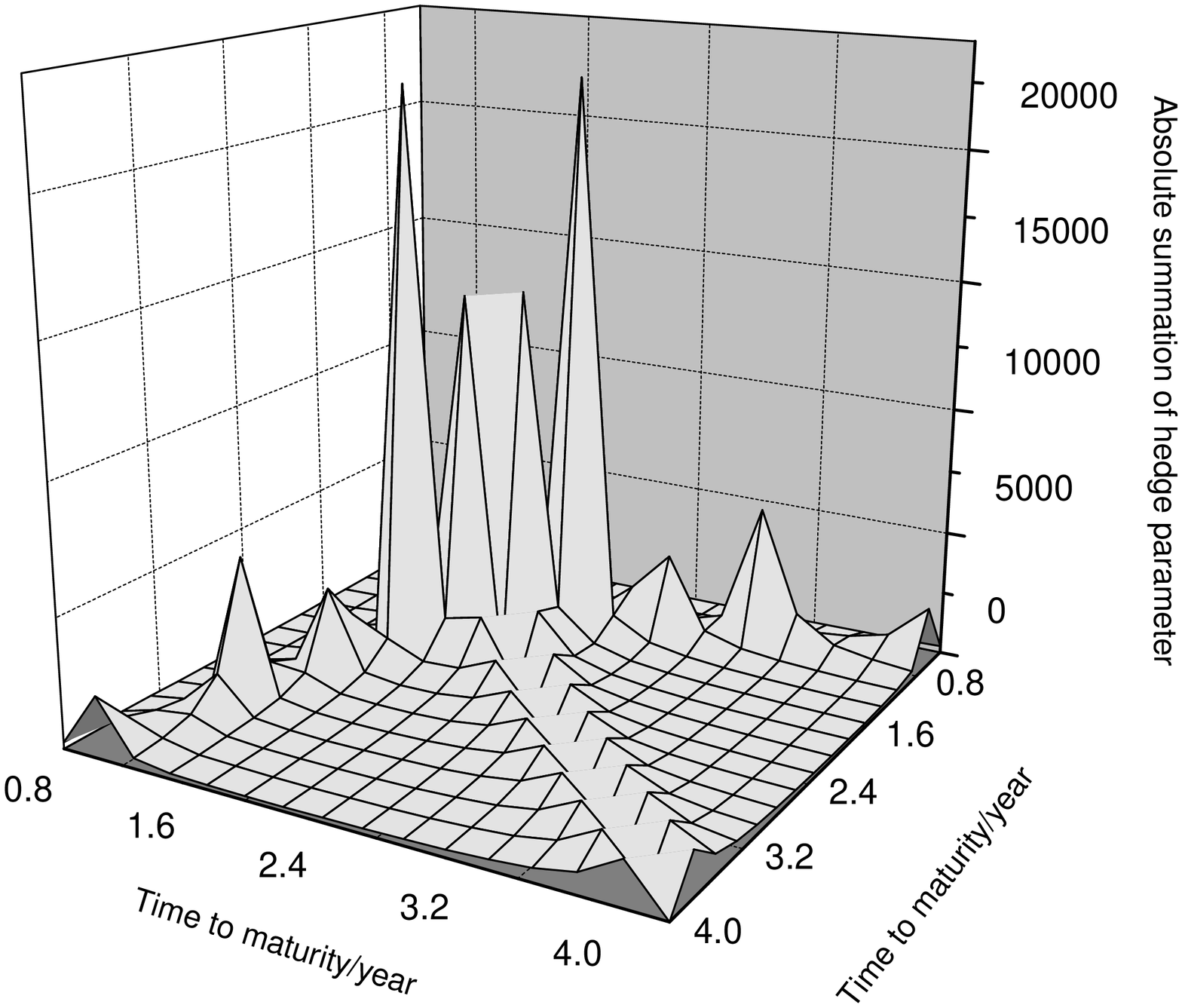, height=12cm ,angle=-90}
  \caption{\emph{Summation of absolute hedge parameters $|n_{1}|+|n_{2}|$ for
  two LIBOR futures, $T_{1}$ and $T_{2}$. The portfolio
  $\Pi(t)=Cap(t,1,4) + \sum_{i=1}^{2}n_{i}(t){\cal F}(t,T_{i})$ involves a
  stochastic hedge against one forward rate with both Delta and Gamma hedging.}}
  \label{fig:deltagamma}
\end{figure}

In Fig. \ref{fig:deltagamma}, we investigate hedging one forward
rate with two LIBOR futures by employing both Delta and Gamma
hedging. The portfolio is given by
\begin{eqnarray}
\Pi(t)=Cap(t,t_{*},T) +\sum_{i=1}^{2}n_{i}(t){\cal F}(t,T_{i}) \,
\nonumber
\end{eqnarray}
where stochastic Delta hedging
$\frac{\partial}{\partial{f_{1}}}E[\Pi(t)|f_{1}]=0$ and stochastic
Gamma hedging
$\frac{\partial^{2}}{\partial{f_{1}^{2}}}E[\Pi(t)|f_{1}]=0$ are
employed.

From the previous case, we can hedge against $f(t,\delta)$ in
order to obtain a minimum portfolio involving the least amount of
short and long positions. The diagonal reports that two LIBOR
futures with the same maturity reduces to Delta hedging with one
LIBOR futures. The data from which Fig. \ref{fig:deltagamma} is
plotted illustrates that selling 38 contracts of $L(t,t+6\delta)$ and
buying 71 $L(t,t+\delta)$ contracts identifies the minimum
portfolio. More explicitly, the variables in the portfolio are
given as

\begin{tabular}{|c|c|c|c|c|}
  \hline
  $T_{1}$ & $T_{2}$ & $x_{h1}$ & $n_{1}$ & $n_{2}$ \\
  \hline
  1.5 year & 0.25 year & 0.25 year  & -38 & 71 \\
  \hline
\end{tabular}

\subsubsection{Hedging Against Two Forward Rates with Two LIBOR
Futures}

In addition, we consider hedging fluctuations in two forward
rates. Specifically, we study a portfolio comprised of two LIBOR
futures and one Caplet $\Pi(t)=Cap(t,t_{*},T)
+\sum_{i=1}^{2}n_{i}(t){\cal F}(t,T_{i})$ where the parameters
$n_{i}$ are fixed by Delta hedging
$\frac{\partial}{\partial{f_{1}}}E[\Pi(t)|f_{1},f_{2}]=0$ and
and Cross Gamma hedging $\frac{\partial^{2}}{\partial{f_{1}}\partial{f_{2}}}E[\Pi(t)|f_{1},f_{2}]=0$.

\begin{figure}[h]
  \centering
  \epsfig{file=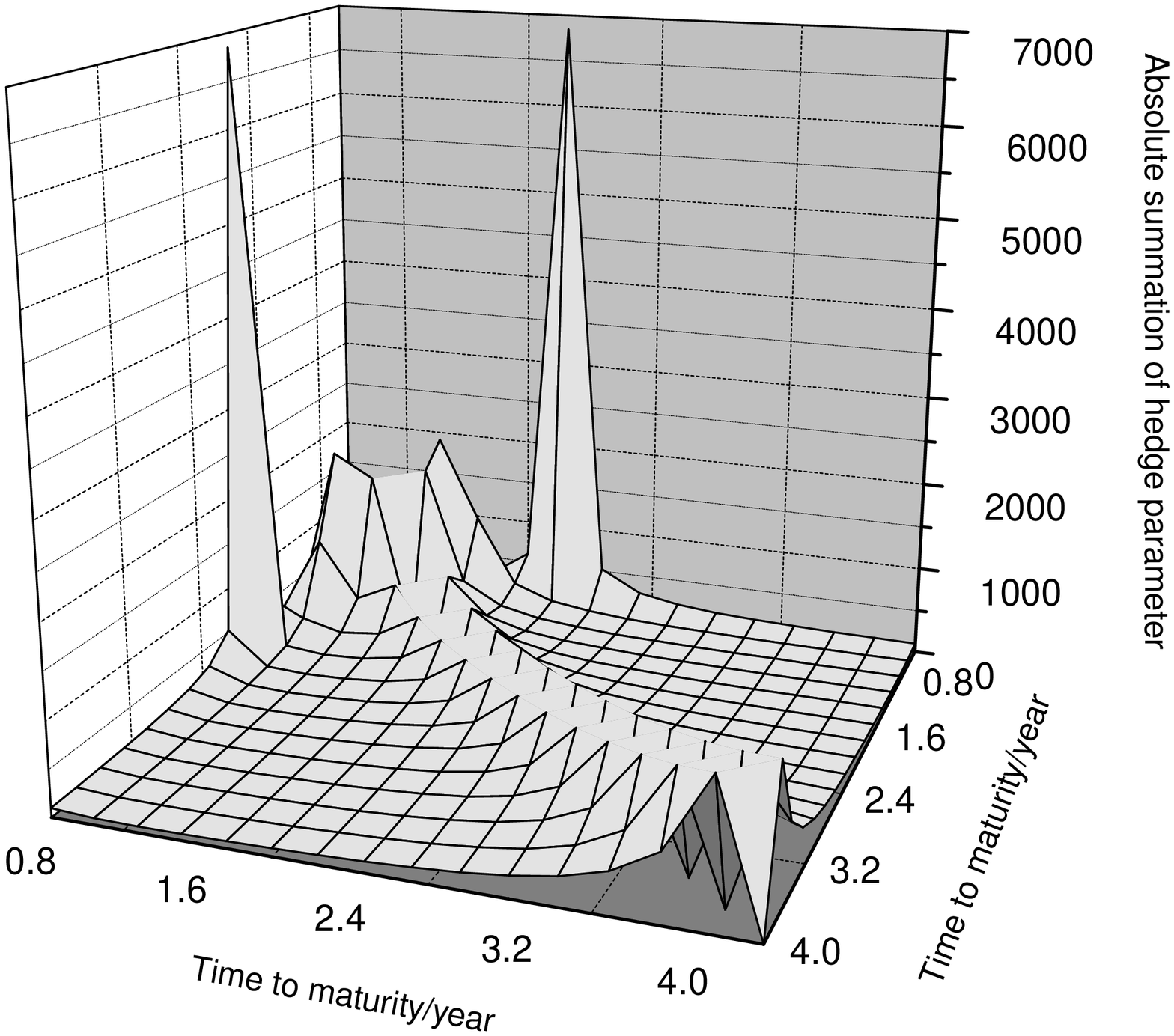, height=12cm ,angle=-90}
  \caption{\emph{Summation of absolute hedge parameters $|n_{1}|+|n_{2}|$ for two LIBOR futures maturities, $T_{1}$ and $T_{2}$.
  The portfolio $\Pi(t)=Cap(t,1,4) + \sum_{i=1}^{2}n_{i}(t){\cal F}(t,T_{i})$ involves stochastic hedging against two forward rates,
   with both Delta and Cross Gamma hedging.} }
  \label{fig:deltacrossgamma}
\end{figure}

The result is displayed in Fig. \ref{fig:deltacrossgamma} where we
hedge against two short maturity forward rates, such as
$f(t,\delta)$ and $f(t,2 \delta)$. Again the data from which Fig.
\ref{fig:deltacrossgamma} is plotted illustrates that buying 45 contracts
of $L(t,t+15\delta)$ and selling 25 $L(t,t+3\delta)$ contracts
forms the minimum portfolio. More explicitly, the variables in the
portfolio are given as

\begin{tabular}{|c|c|c|c|c|c|}
  \hline
  $T_{1}$ & $T_{2}$ & $x_{h1}$ & $x_{h2}$ & $n_{1}$ & $n_{2}$ \\
  \hline
  3.75 year & 0.75 year & 0.25 year & 0.5 year & 45 & -25 \\
  \hline
\end{tabular}

Fig. \ref{fig:deltagamma} and Fig. \ref{fig:deltacrossgamma}
result from summing the absolute values of the hedge parameters
(as in equation (\ref{portfoliozero})) which depend on the
maturities of the LIBOR futures $T_{i}$. The corresponding
empirical results are consistent with our earlier
discussion.\footnote {If we choose the hedge portfolio by
minimizing $\sum_{i=1}^{N}n_{i}$, we find that the minimum
portfolio requires 1500 contracts (long the short maturity and
short their long maturity counterparts).}

\section{Conclusion}

LIBOR-based Caps and Floors are important financial instruments for
managing interest rate risk. However, the multiple payoffs
underlying these contracts complicates their pricing as the LIBOR
term structure dynamics are not perfectly correlated. A field theory
model which allows for imperfect correlation between every LIBOR
maturity overcomes this difficulty while maintaining model
parsimony.

Furthermore, hedge parameters for the field theory model are
provided for risk management applications. Although the field theory
model implies an incomplete market since hedging cannot be conducted
with an infinite number of interest rate dependent securities in
practice, the correlation structure between LIBOR rates is exploited
to minimize risk. An empirical illustration demonstrates the implementation of our
model.

\section{Acknowledgment}

The data in our empirical tests was generously provided by
Jean-Philippe Bouchaud of Science and Finance, and consists of daily
closing prices for quarterly Eurodollar futures contracts as
described in Bouchaud, Sagna, Cont, El-Karoui, and Potters
\cite{data} as well as Bouchaud and Matacz \cite{data1}.

\appendix

\section{Residual Variance}
First, consider the variance of a Cap in the field theory model.
Define the Delta of the Cap, $\frac{\partial Cap(t,T)}{\partial
\int_{T}^{T+\ell} f(t,x) dx }$, as $\chi$
\begin{eqnarray}
\chi &\equiv & - V B(t,T)
\int^{+\infty}_{-\infty}\frac{d{G}}{\sqrt{2\pi{q}^2}}\frac{1}{q^2}
\left( G-\int_{T}^{T+\ell}dx{f}(t,x)-\frac{q^2}{2}
\right)\nonumber\\
&&\times\left\{e^{-\frac{1}{2q^2} \left(
G-\int_{T}^{T+\ell}dx{f}(t,x) -\frac{q^2}{2}
\right)^2}(X-e^{-G})_{+}\right\} \, .\label{chi_define}
\end{eqnarray}
The result in equation (\ref{eqncap}) for the Cap price implies
that
\begin{eqnarray}
\frac{d{Cap(t,T)}}{d{t}}&=& - V B(t,T) \int^{+\infty}_{-\infty}\frac{d{G}}{\sqrt{2\pi{q}^2}}\frac{1}{q^2} \left( G-\int_{T}^{T+\ell}dx{f}(t,x)-\frac{q^2}{2} \right) \int_{T}^{T+\ell}\frac{\partial{f(t,x)}}{\partial{t}}dx\nonumber\\
&&\times\left\{e^{-\frac{1}{2q^2} \left( G-\int_{T}^{T+\ell}dx{f}(t,x)-\frac{q^2}{2} \right)^2} \left(X-e^{-G} \right)_{+}\right\}\\
&=&\chi \left( \int_{T}^{T+\ell}\frac{\partial{f}(t,x)}{\partial{t}}dx \right) \\
&=&\chi \left(
\int_{T}^{T+\ell}dx\alpha(t,x)+\int_{T}^{T+\ell}dx\sigma(t,x)A(t,x)
\right)
\end{eqnarray}
with $E\left[ \frac{dCap(t,T)}{dt} \right] = \left(
\int_T^{T+\ell} dx \alpha(t, x) \right) dt$ since $E[A(t, x)] =
0$. Therefore, the resulting variance equals
\begin{equation}
\frac{dCap(t,T)}{\epsilon} - E\left[ \frac{dCap(t,T)}{ \epsilon }
\right] =\chi\int_T^{T+\ell} dx \sigma(t,x) A(t, x) \, .
\end{equation}
With $\delta(\cdot) = \frac{1}{\epsilon}$ representing a delta
function, squaring this expression and invoking the property that
$E[A(t, x)A(t, x')] = \delta(0) D(x,x';t) = \frac{D(x,x';t)}{dt}$
results in the instantaneous Cap price variance being
\begin{equation}
Var \left[ \frac{dCap(t,T)}{\epsilon} \right] = \frac{1}{\epsilon}
\chi^2 \int_T^{T+\ell} dx \int_T^{T+\ell} dx' \sigma(t, x)
D(x,x';t) \sigma(t, x') \, .
\end{equation}
The quantity $\epsilon$ signifies a small step forward in time.
The underlying intuition is that we are converting a portfolio of
futures contracts to one involving another function of LIBOR
rates. Then, the instantaneous variance of a LIBOR portfolio is
considered. For a LIBOR portfolio, $\hat{\Pi}(t) =
V\ell\sum_{i=1}^N \Delta_i L(t,T_i)$, the following result holds,
\begin{equation}
\frac{d\hat{\Pi}(t)}{dt} - E \left[ \frac{d\hat{\Pi}(t)}{dt}
\right] = \sum_{i=1}^N \Delta_i \hat{L}(t, T_i)
\int_{T_i}^{T_i+\ell} dx \sigma(t, x) A(t, x)
\end{equation}
where $\hat{L}(t,T_i) = V e^{\int_{T_i}^{T_i+\ell}f(t,x)dx} =
\frac{V}{f(t,T_i,T_i+\ell)}$ and
\begin{equation}
Var \left[ \frac{d\hat{\Pi}(t)}{dt} \right]
=\frac{1}{\epsilon}\sum_{i=1}^N \sum_{j=1}^N \Delta_i \Delta_j
\hat{L}(t,T_i) \hat{L}(t,T_j) \int_{T_i}^{T_i+\ell} dx
\int_{T_j}^{T_j+\ell} dx \sigma(t, x) D(x,x';t) \sigma(t, x') \,
.\label{eq:liborvariance}
\end{equation}
The (residual) variance of the hedged portfolio $$\Pi(t) =
Cap(t,T) + \sum_{i=1}^N \Delta_i {\cal F}(t,T_i)$$ is then
computed in a straightforward manner. Equation
(\ref{eq:liborvariance}) implies the hedged portfolio's variance
equals
\begin{equation}
  \begin{split}
    \label{eq:hedgingcap}
    &\chi^2 \int_T^{T+\ell} dx \int_T^{T+\ell} dx' \sigma(t, x) \sigma
    (t, x') D(x,x';t) \\
    +& 2 \chi \sum_{i=1}^N \Delta_i \hat{L}(t,T_i) \int_T^{T+\ell} dx \int_{T_i}^{T_i+\ell} dx'
    \sigma(t, x) \sigma (t, x') D(x,x';t)  \\
    +& \sum_{i=1}^N \sum_{j=1}^N \Delta_i \Delta_j \hat{L}(t,T_i) \hat{L}(t,T_j)
    \int_{T_i}^{T_i+\ell} dx     \int_{T_j}^{T_j+\ell} dx' \sigma(t, x) \sigma(t, x') D(x,x';t) \, .
  \end{split}
\end{equation}

\section{Conditional Probability of Hedging One Forward Rate}

Using the results of the Gaussian models in Baaquie
\cite{baaqbook}, after a straightforward but tedious calculation,
the following is derived from equations (\ref{condcap}) and (\ref{condlibor})
\begin{eqnarray}
\Psi(G|f_{h})&=&\frac{\chi}{\sqrt{2\pi{Q}^{2}}}\exp\left[-\frac{1}{2Q^2}(G-G_{0})^{2}\right]\\
\Phi(G|f;t_{h},T_{n1})&=&\frac{1}{\sqrt{2\pi{Q}_{1}^{2}}}\exp\left[-\frac{1}{2Q_{1}^2}(G-G_{1})^2\right] \, .
\end{eqnarray}
The notations are shown as follow
\begin{eqnarray}
X&=&\frac{1}{1+\ell k}\,\,\,;\,\,\,\tilde{V}=(1+\ell k)V\nonumber\\
\chi&=&\exp\left\{-\int_{t_{h}}^{T_{n}}dx{f}(t_{0},x)-\int_{M_{1}}\alpha(t,x)+\frac{1}{2}E+\frac{C}{A}(f(t_{0},x_{h})+\int_{t_{0}}^{t_{h}}dt\alpha(t,x_{h})-f-\frac{C}{2})\right\}\nonumber\\
d_{+}&=&(\ln{x}+G_{0})/Q \qquad ;\qquad d_{-}=(\ln{x}+G_{0}-Q^{2})/Q\nonumber\\
G_{0}&=&\int_{T_{n}}^{T_{n}+\ell}dx{f}(t_{0},x)-F-\frac{B}{A}(f(t_{0},x_{h})-C-f+\int_{t_{0}}^{t_{h}}dt\alpha(t,x_{h}))+\frac{q^{2}}{2}\nonumber\\
Q^{2}&=&q^{2}-\frac{B^{2}}{A}\nonumber\\
G_{1}&=&\int_{T_{n1}}^{T_{n1}+\ell}dx{f}(t_{0},x)+\int_{M_{3}}\alpha(t,x)-\frac{B_{1}}{A}
(f(t_{0},x_{h})-\int_{t_{0}}^{t_{h}}dt\alpha(t,x_{h})-f)\nonumber\\
Q_{1}^{2}&=&D-\frac{B_{1}^{2}}{A}\nonumber\\
A&=&\int_{t_{0}}^{t_{h}}dt\sigma(t,x_{h})^{2}D(t,x_{h},x_{h};T_{FR})\nonumber\\
B&=&\int_{M_{2}}\sigma(t,x_{h})D(t,x_{h},x;T_{FR})\sigma(t,x)\nonumber\\
B_{1}&=&\int_{\tilde{M}_{1}}\sigma(t,x_{h})D(t,x_{h},x;T_{FR})\sigma(t,x)\nonumber\\
C&=&\int_{M_{1}}\sigma(t,x_{h})D(t,x_{h},x;T_{FR})\sigma(t,x)\nonumber\\
D&=&\int_{\tilde{{\cal
Q}}_{1}}\sigma(t,x)D(t,x,x^{'};T_{FR})\sigma(t,x^{'})\nonumber\\
q^{2}&=&\int_{{\cal Q}_{2}+{\cal
Q}_{4}}\sigma(t,x)D(t,x,x^{'};T_{FR})\sigma(t,x^{'})\nonumber \\
E&=&\int_{{\cal Q}_{1}}\sigma(t,x)D(t,x,x^{'};T_{FR})\sigma(t,x^{'})\nonumber\\
F&=&\int_{t_{0}}^{t_{h}}dt\int_{t_{h}}^{T_{n}}dx\int_{T_{n}}^{T_{n}+\ell}dx^{'}
\sigma(t,x)D(t,x,x^{'};T_{FR})\sigma(t,x^{'}) \, . \nonumber
\end{eqnarray}
The domain of integration is given in Figs. \ref{fig1} and
\ref{fig2}. It can be seen that the unconditional probability
distribution for the Cap and LIBOR futures yields volatilities
$q^2$ and $D$ respectively. Hence the conditional expectation
reduces the volatility of Cap by $\frac{B^{2}}{A}$, and by
$\frac{B_{1}^{2}}{A}$ for the LIBOR futures. This result is
expected since the constraint imposed by the requirement of a
conditional probability reduces the allowed fluctuations of the
instruments.

It could be the case that there is a special maturity time $x_{h}$
which causes the largest reduction in conditional variance. The
answer is found by minimizing the conditional variance
\begin{eqnarray}
\tilde{Cap}(t_{h},t_{*},T_{n};f_{h})&=&\chi\tilde{V}(xN(d_{+})-e^{-G_{0}+\frac{Q^{2}}{2}}N(d_{-}))\\
\tilde{L}(t_{h},T_{n1};f_{h})&=&e^{G_{1}+\frac{Q_{1}^{2}}{2}} \, . \label{condipro}
\end{eqnarray}
Recall the hedging parameter is given by equation (\ref{parameter}).
Using equation (\ref{condipro}) and setting $t_{0}=t$, $t_{h}=t+\epsilon$,
we get an (instantaneous) stochastic Delta hedge parameter $\eta_1(t)$ equal to
\begin{eqnarray}\label{hedgweight}
&& \frac{C\cdot{\tilde{Cap}}(t,t_{*},T_{n};f_{h})-B\cdot\chi\cdot\tilde{V}\cdot \left[ xN^{'}(d_{+})/Q+e^{-G_{0}+\frac{Q^2}{2}}N(d_{-})-e^{-G_{0}+\frac{Q^2}{2}}N^{'}(d_{-})/Q \right]}{e^{G_{1}+\frac{Q_{1}^2}{2}}\cdot{B}_{1}} \, .
\end{eqnarray}

\section{HJM Limit of Hedging Function}

The HJM-limit of the hedging functions is analyzed for the
specific exponential function considered by Jarrow and Turnbull \cite{JTtext}
\begin{eqnarray}
\sigma_{hjm}(t,x)=\sigma_{0}e^{\beta(x-t)} \, ,  \label{HJMvol}
\end{eqnarray}
which sets the propagator $D(t,x,x^{'};T_{FR})$ equal to one. It can be shown that
\begin{eqnarray}
A&=&\frac{\sigma_{0}^{2}}{2\beta}e^{-2\beta{x}_{h}}(e^{2\beta{t}_{h}}-e^{2\beta{t}_{0}})\nonumber\\
B&=&\frac{\sigma_{0}^{2}}{2\beta^{2}}e^{-\beta{x}_{h}}(e^{-\beta{T}_{n}}-e^{-\beta{T}_{n}+\ell})(e^{2\beta{t}_{h}}-e^{2\beta{t}_{0}})\nonumber\\
B_{1}&=&\frac{\sigma_{0}^{2}}{2\beta^{2}}e^{-\beta{x}_{h}}(e^{-\beta{T}_{n1}}-e^{-\beta{T}_{n1}+\ell})(e^{2\beta{t}_{h}}-e^{2\beta{t}_{0}})\nonumber \\
C&=&\frac{\sigma_{0}^{2}}{2\beta^{2}}e^{-\beta{x}_{h}}(e^{-\beta{t}_{h}}-e^{-\beta{T}_{n}})(e^{2\beta{t}_{h}}-e^{2\beta{t}_{0}})\nonumber\\
D&=&\frac{\sigma_{0}^{2}}{2\beta^{3}}(e^{-\beta{T}_{n1}+\ell}-e^{-\beta{T}_{n1}})^2(e^{2\beta{t}_{h}}-e^{2\beta{t}_{0}})\nonumber\\
E&=&\frac{\sigma_{0}^{2}}{2\beta^{3}}(e^{-\beta{T}_{n}}-e^{-\beta{t}_{h}})^2(e^{2\beta{t}_{h}}-e^{2\beta{t}_{0}})\nonumber\\
F&=&\frac{\sigma_{0}^{2}}{2\beta^{3}}(e^{-\beta{T}_{n}+\ell}-e^{-\beta{T}_{n}})(e^{-\beta{T}_{n}}-e^{-\beta{t}_{h}})(e^{2\beta{t}_{h}}-e^{2\beta{t}_{0}})\, . \nonumber
\end{eqnarray}
The exponential volatility function given in equation (\ref{HJMvol})
has the remarkable property, similar to the case found for the
hedging of Treasury Bonds in Baaquie \cite{baaqbook}, that
\begin{eqnarray}
Q_{1}^{2}(hjm)=D_{hjm}-\frac{B^{2}_{1 hjm}}{A_{hjm}}\equiv 0 \, .
\end{eqnarray}
Hence, the conditional probability for the LIBOR futures is
deterministic. Indeed, once the forward rate $f_{h}$ is fixed,
the following identity is valid
\begin{eqnarray}
\tilde{L}_{hjm}(t_{h},T_{n1};f_{h}) & \equiv& L(t_{h},T_{n1}) \, .
\end{eqnarray}
In other words, for the volatility function in equation (\ref{HJMvol}),
the LIBOR futures for the HJM model is exactly determined by one of
the forward rates.

However, the conditional probability for the Cap is not deterministic
since the volatility from $t_{h}$ to $t_{*}$, before the Cap's
expiration, is not compensated for by fixing the forward rate.

\section{Conditional Probability of Hedging Two Forward Rates}

\begin{figure}[h]
  \centering
  \epsfig{file=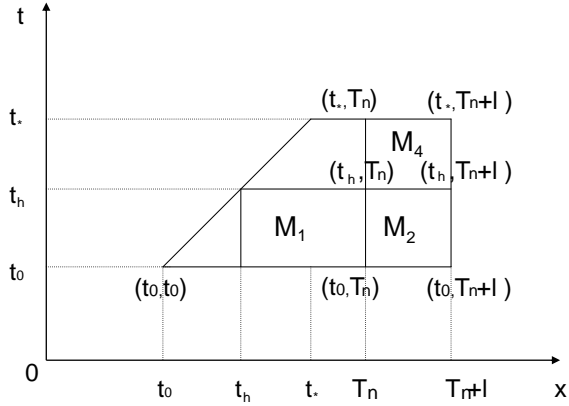, height=14cm, angle=-90}
  \caption{\emph{Domain of integration $M_{1}$, $M_{2}$ and integration cube ${\cal Q}_{1}$, ${\cal Q}_{2}$, ${\cal Q}_{4}$
  where the $x^{'}$ axis has the same limit as its corresponding $x$ axis.}}
  \label{fig1}
\end{figure}

\begin{figure}[h]
  \centering
  \epsfig{file=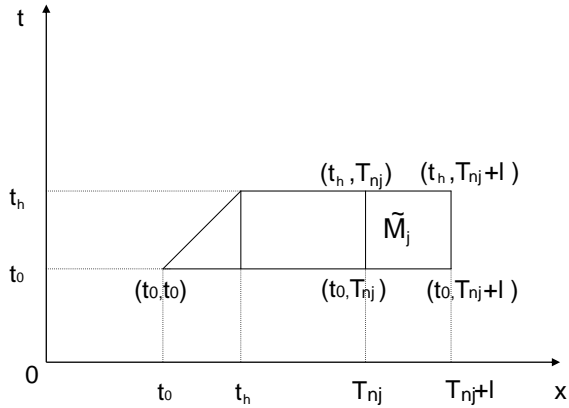, height=14cm, angle=-90}
  \caption{\emph{Domain of integration $\tilde{M}_{j}$ and integration cube $\tilde{{\cal Q}}_{j}$ where the $x^{'}$ axis has the same limit as its corresponding $x$ axis.}}
  \label{fig2}
\end{figure}

When hedging against two forward rates, equations (\ref{condcap}) and (\ref{condlibor}) imply we
have the conditional probability of a Cap given by
\begin{eqnarray}
\Psi(G|f_{1},f_{2})=\frac{\int_{-\infty}^{\infty}\frac{dp}{2\pi}
e^{-\frac{q_{h}^{2}}{2}p^{2}}e^{ip(G-\frac{q_{h}^{2}}{2})}\int{D}f
e^{-\int_{t_{h}}^{T_{n}}f(t_{h},x)}e^{ip\int_{T_{n}}^{T_{n}+l}dx
f(t_{h},x)}\prod_{i=1}^{2}\delta(f({t_{h},x_{i}})-f_{i})e^{S}}{\int{D}f\prod_{i=1}^{2}\delta(f({t_{h},x_{i}})-f_{i})e^{S}} \, ,
\end{eqnarray}
and the conditional probability of LIBOR being
\begin{eqnarray}
\Phi(G|f_{1},f_{2},T_{nj})=\frac{\int{D}f\delta(G-\int_{T_{nj}}^{T_{nj}+\ell}f(t_{h},x)dx)\prod_{i=1}^{2}\delta(f(t_{h},x_{i})-f_{i})e^{S}}
{\int{D}f\prod_{i=1}^{2}\delta(f(t_{h},x_{i})-f_{i})e^{S}} \quad
j=1,2
\end{eqnarray}
which yields
\begin{eqnarray}
\Psi(G|f_{1},f_{2})&=&\frac{\chi}{\sqrt{2\pi{Q}^{2}}}\exp\left[-\frac{1}{2Q^2}(G-G_{0})^{2}\right]\\
\Phi(G|f_{1},f_{2},T_{nj})&=&\frac{1}{\sqrt{2\pi{\tilde{Q}}_{j}^{2}}}\exp\left[-\frac{1}{2\tilde{Q}_{j}^2}(G-\tilde{G}_{j})^2\right]\quad
j=1,2 \,
\end{eqnarray}
under the following notation
\begin{eqnarray}
X&=&\frac{1}{1+\ell k}\,\,\,;\,\,\,\tilde{V}=(1+\ell
k)V\nonumber\\
\chi&=&\exp\left\{-\int_{t_{h}}^{T_{n}}dx{f}(t_{0},x)-\int_{M_{1}}\alpha(t,x)+\frac{1}{2}E+\frac{C_{12}}{\tilde{A}_{12}}(R_{12}-\frac{C_{12}}{2})\right\}\nonumber
\end{eqnarray}
\begin{eqnarray}
d_{+}&=&(\ln{x}+G_{0})/Q \qquad ;\qquad
d_{-}=(\ln{x}+G_{0}-Q^{2})/Q\nonumber\\
G_{0}&=&\int_{T_{n}}^{T_{n}+\ell}dx{f}(t_{0},x)-F-\frac{B_{12}}{\tilde{A}_{12}}(R_{12}-C_{12})+\frac{q^{2}}{2}\nonumber\\
Q^{2}&=&q^{2}-\frac{B_{12}^{2}}{\tilde{A}_{12}}\nonumber\\
\tilde{G}_{j}&=&\int_{T_{nj}}^{T_{nj}+\ell}dx{f}(t_{0},x)+\int_{\tilde{M}_{j}}\alpha(t,x)-\frac{\tilde{B}_{12j}}{\tilde{A}_{12}}
R_{12}\quad j=1,2\nonumber\\
\tilde{Q}_{j}^{2}&=&D_{j}-\frac{\tilde{B}_{12j}^{2}}{\tilde{A}_{12}}\quad j=1,2\nonumber\\
R_{i}&=&f(t_{0},x_{i})+\int_{t_{0}}^{t_{h}}dt\alpha(t,x_{i})-f_{i}\quad
i=1,2\nonumber\\
R_{12}&=&R_{1}-\frac{A_{12}}{A_{2}}R_{2}\nonumber\\
A_{i}&=&\int_{t_{0}}^{t_{h}}dt\sigma(t,x_{i})^{2}D(t,x_{i},x_{i};T_{FR})\quad i=1,2\nonumber\\
A_{12}&=&\int_{t_{0}}^{t_{h}}dt\sigma(t,x_{1})D(t,x_{1},x_{2};T_{FR})\sigma(t,x_{2})\nonumber\\
\tilde{A}_{12}&=&A_{1}-\frac{A_{12}}{A_{2}}\nonumber\\
B_{i}&=&\int_{M_{2}}\sigma(t,x_{i})D(t,x_{i},x;T_{FR})\sigma(t,x)\quad i=1,2\nonumber\\
B_{12}&=&B_{1}-\frac{A_{12}}{A_{2}}B_{2}\nonumber\\
\tilde{B}_{ij}&=&\int_{\tilde{M}_{j}}\sigma(t,x_{i})D(t,x_{i},x;T_{FR})\sigma(t,x)\quad i=1,2;\quad j=1,2\nonumber\\
\tilde{B}_{12j}&=&\tilde{B}_{1j}-\frac{A_{12}}{A_{2}}\tilde{B}_{2j}\quad
j=1,2\ldots, 5\nonumber\\
C_{i}&=&\int_{M_{1}}\sigma(t,x_{i})D(t,x_{i},x;T_{FR})\sigma(t,x)\quad i=1,2\nonumber\\
C_{12}&=&C_{1}-\frac{A_{12}}{A_{2}}C_{2}\nonumber\\
D_{j}&=&\int_{\tilde{{\cal Q}_{j}}}\sigma(t,x)D(t,x,x^{'};T_{FR})\sigma(t,x^{'})\quad j=1,2\nonumber\\
q^{2}&=&\int_{{\cal Q}_{2}+{\cal Q}_{4}}\sigma(t,x)D(t,x,x^{'};T_{FR})\sigma(t,x^{'})\nonumber\\
E&=&\int_{{\cal Q}_{1}}\sigma(t,x)D(t,x,x^{'};T_{FR})\sigma(t,x^{'})\nonumber\\
F&=&\int_{t_{0}}^{t_{h}}dt\int_{t_{h}}^{T_{n}}dx\int_{T_{n}}^{T_{n}+\ell}dx^{'}
\sigma(t,x)D(t,x,x^{'};T_{FR})\sigma(t,x^{'}) \, .
\end{eqnarray}
The domain of integration is given in Figs \ref{fig1} and \ref{fig2}.

Furthermore, an $N$-fold constraint on the instruments would further reduce the variance of the instruments
\begin{eqnarray}
\tilde{Cap}(t_{h},t_{*},T_{n};f_{1},f_{2})&=&\chi\tilde{V}(xN(d_{+})-e^{-G_{0}+\frac{Q^{2}}{2}}N(d_{-}))\\
\tilde{L}(t_{h},T_{nj};f_{1},f_{2})&=&e^{\tilde{G}_{j}+\frac{\tilde{Q}_{j}^{2}}{2}} \, . \label{condipro2}
\end{eqnarray}


\begin{thebibliography}{100}

\bibitem{baaqbook}
B. E. Baaquie, Quantum Finance, {\it Cambridge University Press} (2004).

\bibitem{baaqmar2}
B. E. Baaquie, M. Srikant and M. Warachka, A Quantum Field Theory
Term Structure Model Applied to Hedging, {\it International Journal
of Theoretical and Applied Finance} {\bf 6} (2003) 443-468.

\bibitem{baaqbou}
B. E. Baaquie and J. P. Bouchaud, Stiff Interest Rate Model and
Psychological Future Time, {\it Wilmott Magazine} (April. 2004).

\bibitem{baalibor}
B. E. Baaquie, A Common Market Measure for LIBOR and Pricing Caps,
Floors and Swaps in a Field Theory of Forward Interest Rates, {\it
International Journal of Theoretical and Applied Finance} Vol 8.
No. 8 (2005) 999-1018.

\bibitem{capswap}
Paul Glasserman and Nicolas Merener, Cap and Swaption
Approximation in LIBOR Market Models with Jumps, {\it Journal of
Computational Finance} (2003)

\bibitem{farshid}
Farshid Jamshidian, LIBOR and swap market models and measures,
{\it Finance and Stochastics} {\bf V1 4} (1997)

\bibitem{BGM}
A. Brace., D. Gatarek and M. Musiela, The Market Model of Interest Rate Dynamics, {\it Mathematical Finance} {\bf 9} (1997) 127-155.

\bibitem{data}
J. P. Bouchaud, N. Sagna, R. Cont, N. El-Karoui and M. Potters, Phenomenology of the Interest Rate Curve, {\it Applied Financial
Mathematics}  {\bf 6} (1999) 209-232.

\bibitem{data1}
J. P. Bouchaud and A. Matacz, An Empirical Investigation of the Forward Interest Rate Term Structure, {\it International Journal of Theoretical and Applied Finance} {\bf 3} (2000) 703-729.

\bibitem{HJM}
D. Heath, R. Jarrow and A. Morton, Bond Pricing and the Term Structure of Interest Rates: A New Methodology for Pricing Contingent Claims,
{\it Econometrica} {\bf 60} (1992) 77-105.

\bibitem{JTtext}
R. Jarrow and S. Turnbull, {\it Derivative Securities, Second Edition}, South-Western College Publishing (2000).

\bibitem{MR}
M. Musiela and M. Rutkowski, {\it Martingale Methods in Financial Modeling}, Springer - Verlag {\bf 36} (1997).

\end{thebibliography}
\end{document}